\documentclass[prl, twocolumn, showpacs]{revtex4}
\usepackage{amsfonts}
\usepackage{graphicx}
\newcommand{\bee}{\begin{eqnarray}}
\newcommand{\eend}{\end{eqnarray}}
\newcommand{\rmd}{{\rm d}}
\newcommand{\rme}{{\rm e}}

\begin{document}
%%%%%%%%%%%%%%%%%%%%%%%%%%%%%%%%%%%%%%%%%%%%%%%%%%%%%%%%%%%%%%%%%%%%%
\title{Modified Coulomb law in a strongly magnetized vacuum
}
\author{Anatoly E. Shabad$^1$ and Vladimir V. Usov$^2$}
\affiliation{ $^1$ P.N. Lebedev Physics Institute, Moscow 117924,
Russia\\
$^2$Center for Astrophysics, Weizmann Institute, Rehovot 76100,
Israel}

%%%%%%%%%%%%%%%%%%%%%%%%%%%%%%%%%%%%%%%%%%%%%%%%%%%%%%%%%%%%%%%%%%%%%
\begin{abstract}
We study  electric potential of a charge placed in a strong
magnetic field $B\gg B_0=4.4\cdot 10^{13}$G, as modified by the
vacuum polarization. In such field the electron Larmour radius is
much less than its Compton length. At the Larmour distances
  a scaling law occurs, with the potential
determined by a magnetic-field-independent function. The scaling
regime implies  short-range interaction, expressed by Yukawa law.
The electromagnetic interaction regains its long-range character
at distances larger than the Compton length, the potential
decreasing across $\bf B$ faster than along. Correction to the
nonrelativistic ground-state energy of a hydrogenlike atom is
found. In the limit $B=\infty$, the modified potential becomes the
Dirac $\delta$-function plus a regular background. With this
potential the ground-state energy is finite - the best pronounced
effect of the vacuum polarization.
\end{abstract}
 \maketitle
There is now
compelling evidence that many compact astronomical objects (soft
gamma-ray repeaters, anomalous X-ray pulsars, and some radio
pulsars) identified with neutron stars have surface magnetic
fields as high as $\sim 10^{14}-10^{15}$~G \cite{TD95}. More
strong magnetic fields ($B\sim 10^{16}-10^{17}$~G) are predicted
to exist at the surface of cosmological gamma-ray bursters if they
are rotation-powered neutron stars similar to radio pulsars
\cite{U92}. All these fields, however, are much smaller than the
maximum value inherent in quantum electrodynamics \cite{prl}.

 Vacuum in an external magnetic field
 $B$ behaves as an anisotropic dielectric medium with spatial and
frequency dispersion, ({\em e.g.}, \cite{shabtrudy}). These
properties may become important, provided that the field strength
achieves the characteristic value $B_0=m^2/e\simeq 4.4\times
10^{13}$~G, where $m$ is the electron mass and $e$ is its charge.
[Henceforth, we set $\hbar=c=1$ and refer to the Heaviside-Lorentz
system of units.] Although much work has been devoted  to  study
of electromagnetic wave propagation in the magnetized vacuum,
problems of electro- and magneto-statics in this medium did not
attract sufficient attention, save Refs. \cite{loskutov1,
loskutov2}, where corrections to the Coulomb law were found when
these are small: for $B/B_0 \ll 1$ in \cite{loskutov1}, or at
large distances from the source for $1\ll B/B_0\ll
3\pi\alpha^{-1}$ in \cite{loskutov2} ($\alpha = e^2/4\pi=1/137$).

In this Letter, we find that for sufficiently large $b\equiv B/B_0
\gg 1$ the electric field produced by a pointlike charge at rest
may be significantly modified by the vacuum polarization, the
modification being determined by the characteristic factor $\alpha
b$. The modified Coulomb potential in the close vicinity of its
charge, characterized by the Larmour length $L_{\rm
B}=(eB)^{-1/2}=(1/m\sqrt{b})$, goes steeper than the standard one,
following a Yukawa law, whereas it obeys a long-range "anisotropic
Coulomb law" far from the source, at distances characterized by
the electron Compton length $m^{-1}$, $m^{-1}\gg L_{\rm B}$.
(Details of the corresponding derivations can be found in the
accompanying preprint \cite{SUPRD2}). The short-range part of the
modified potential tends to the Dirac $\delta$-function in the
limit $b\rightarrow\infty$.  The modification of the Coulomb law
should affect, first of all, the field of an atomic nucleus,
placed in a magnetic field. We determine the corresponding
correction to the lowest energy level of a hydrogenlike atom.
Unlike the famous result of Ref. \cite{elliott}, referred to in
many speculations on
 behavior of matter on the surface of strongly magnetized neutron stars
({\it e.g.,} the review \cite{HD06} and references therein), we
find that the (vacuum-polarization-corrected) ground state energy
remains finite in the limit of infinite magnetic field.

Let the constant and homogeneous  magnetic field $\bf B$ be
directed along axis 3 in the  frame where the pointlike
 charge $q$ is at rest in the origin ${\bf x}=\{x_1, x_2, x_3\}$=0,
and no external electric field exists. By using the tensor
decomposition of the photon propagator over eigenmodes in a
magnetic field \cite{batalin, annphys} it is straightforward to
show \cite{SUPRD2} that electrostatic potential $A_0$ produced by
this charge  has the  form
\begin{eqnarray}\label{A0} A_0({\bf
x})=\frac{q}{(2\pi)^3}\int\frac{\exp({-{\rm i}{\bf kx}})\,{\rm
d}^3k}{{\bf k}^2-\kappa_2(k_3^2,k_\perp^2)},\end{eqnarray} while
its vector potential is zero, $A_{1,2,3}({\bf x})=0$. The static
charge gives rise to electric field only, as it might be expected.
Here $\kappa_2$ is one (out of three) eigenvalue(s) of the
polarization operator $\Pi_\mu^{~\nu}$: $\Pi_\mu^{~\nu}
\flat_\nu^{(a)}=\kappa_a\flat_\mu^{(a)}, a=1,2,3, ~\mu,
\nu=0,1,2,3$. The eigenvectors $\flat_\nu^{(a)}$ are 4-potentials
of the eigenmodes. The eigenvalues depend on two combinations of
the photon momentum components $k_3^2-k_0^2$ and
$k_\perp^2=k_1^2+k_2^2$ taken at zero frequency $k_0=0$. Eq.
(\ref{A0}) is approximation-independent and axial symmetric:
 $A_0({\bf x})=A_0(|x_3|,x_\perp)$,
$x_\perp=\sqrt{x_1^2+x_2^2}$.

Eq. (\ref{A0}) indicates that only mode-2 photons mediate
electrostatic interaction. This fact may be better understood, if
we examine electric and magnetic fields intrinsic to the virtual
(off-shell) photons of this mode, obtained from its 4-vector
potential $\flat^{(2)}_\nu=(k_3\; 0\; 0\; k_0)_\nu$. It may be
seen from explicit representation for these fields \cite{annphys}
that in the limit $k_0=0$ the magnetic field of mode-2 photon
disappears, while its electric field is collinear with ${\bf k},$
{\it i.e.}, it becomes a purely longitudinal virtual photon.
Virtual photons of other modes, in the static case, are carriers
of stationary magnetic fields. For instance, mode-1 photons are
responsible for the field produced by a constant current, flowing
parallel to the external magnetic field.

In the asymptotic regime of high magnetic field, $~eB\gg k_3^2~$
and $~B\gg m^2/e\equiv B_0~$, with the accuracy to terms that only
grow with $B$ as its logarithm or slower, the eigenvalue
$\kappa_2$, calculated within the one-loop approximation in
\cite{batalin,
 annphys}, acquires
 the form \cite{loskutov2, kratkie2}
\begin{eqnarray}
\kappa_2(k_3^2,k_\perp^2) =-\frac{2\alpha bm^2}{\pi }\exp \left(-
\frac{k_\bot^2}{2m^2b}
\right)T\left(\frac{k_3^2}{4m^2}\right),\label{2}\end{eqnarray}
\begin{eqnarray}
T(y)=y\int_{0}^1\frac{(1-\eta^2)\rm d \eta}{1+y(1-\eta^2)}.
\label{T}\end{eqnarray} %Here $b =B/B_0$.
Note the properties: $T(y\rightarrow 0)\simeq 2y/3,\;T(\infty)=1$.

The deviation of the potential (\ref{A0}) from the standard
Coulomb potential $A_0^{\rm C}({\bf
x})=q/(4\pi\sqrt{x_\perp^2+x_3^2})$ is
\begin{eqnarray}\label{difference2}\Delta A_0({\bf x})\equiv A_0^{\rm C}({\bf x})-A_0({\bf
x})=\frac q{8\pi^2}\int_0^\infty J_0(k_\perp x_\perp){\rm d}
k_\perp^2 \nonumber\\\hspace{-1.5cm} \times
\int_{-\infty}^\infty\left(\frac{\exp({-{\rm i} k_3x_3})}{
k_\perp^2+k_3^2} -\frac{\exp({-{\rm i} k_3x_3})}{
k_\perp^2+k_3^2-\kappa_2(k_3^2,k_\perp^2)}\right){\rm d} k_3.
\end{eqnarray} Here $J_0$ is the Bessel function of order zero.
This integral defines $\Delta A_0 ({\bf x})$ as a finite function
of the coordinates in the origin, unlike $A_0$ and $A_0^{\rm C}$.
%Integral (\ref{difference2}) with $\kappa_2$ given by eq. (\ref{2}
%is fast converging, and hence the integration over $k_3$ is as a
As the integration variable $k_3$ in (\ref{difference2})
approaches  the large values $\pm \sqrt{eB}$, one has $\kappa_2\ll
k_3^2$. Hence, no essential contribution comes from the
integration over the region $|k_3|>\sqrt{eB}$, wherein eq.
(\ref{2}) is not valid. The expression to be substituted for it
there is even smaller (since it does not contain the large factor
$b$) \cite{footnote}. The leading terms of the expansion of
(\ref{A0}) near the origin $x_3=x_\perp=0$ are
\begin{eqnarray}\label{expand}\hspace{-0.3cm}A_0({\bf x})\simeq \frac q{4\pi}
\left(\frac 1
{|\bf x|}-2mC\right),\;\; C=\frac{2\pi }{q m}\Delta A_0(0)>0,
\end{eqnarray} where $C$ is a  constant depending
on the  magnetic field. %Next terms of the expansion near ${\bf
%x}=0$ are presented in \cite{SUPRD2}.

%%If  $1\ll b\ll 2\pi/\alpha\sim 10^3,$ we may keep only the
%first-order term in $\kappa_2$ in (\ref{difference2}) to derive
%the magnetic analog of the Uehling potential \cite{blp}. At
%$x_\perp =0$ it is \bee\label{uehling2}\Delta A_0(x_3,0)\simeq
%\frac{q\alpha b m}{8\pi^2}\int_0^{\pi/2}
%\rme^\frac{-2m|x_3|}{\cos\phi}\cos^2\phi~\rmd\phi.\eend This
%correction is of the order of $\alpha b$, {\it i.e.,} is much
%larger than $\alpha$. Eq. (\ref{uehling2}) implies $C\simeq\alpha
%b/16$.

More exactly the singular behavior near the origin will be
presented if we note that it is provided by the integration over
large $k_3$ (and $k_\perp$) in (\ref{A0}), where we may set
$T({k}_3^2/4m^2)=T(\infty)=1$ and perform $k_3$-integration in
Eq.(\ref{A0}) by calculating residues. Then

%It is remarkable that in the infinite-magnetic-field limit,
%potential (\ref{A0}), measured in the inverse Larmour length
%$L_{\rm B}^{-1}=\sqrt{eB}$ units, becomes a universal,
%magnetic-field-independent function of coordinates, measured in
%Larmour units $L_{\rm B}$. To establish this scaling regime,
%change the variables in the integral (\ref{A0})
%$\widetilde{k}_i={k}_iL_{\rm B},$ $i=1,2,3$ and define the new
%dimensionless coordinates
%\bee\label{coord}\widetilde{x}_3=x_3/L_{\rm B},\qquad
%\widetilde{x}_\perp =x_\perp /L_{\rm B}.\eend Unless $\tilde{k}_3$
%is too small (small $\tilde{k}_3$ become most essential for large
%$\tilde{x}_3$), we set $T(\widetilde{k}_3^2/4m^2L_{\rm
%B}^2)=T(\infty)=1$ and perform $k_3$-integration in Eq.(\ref{A0})
%by calculating residues to get
\begin{eqnarray}\label{universal}A_0({\bf x})
\simeq \frac{\widetilde{A}_0(\widetilde{\bf x})}{L_{\rm B}} =\frac
q{4\pi L_{\rm B}}\int_0^\infty J_0(\widetilde{k}_\perp\widetilde{
x}_\perp)\widetilde{k}_\perp\nonumber\\
\times \frac{\exp\left[-|\widetilde{x}_3|
\sqrt{\widetilde{k}_\perp^2
+({2\alpha}/\pi)\exp\left(-\widetilde{k}_\perp^2/
2\right)}\;\right]} {\sqrt{\widetilde{k}_\perp^2+(2\alpha/\pi)\exp
\left(-\widetilde{k}^2_\perp/2\right)}} {\rm d}
\widetilde{k}_\perp.\end{eqnarray} Here
$\widetilde{A}_0(\widetilde{\bf x})$ is a dimensionless,
external-field-independent function of the arguments
$\widetilde{x}_3=x_3/L_{\rm B},$ $\widetilde{x}_\perp =x_\perp
/L_{\rm B}$, and the integration variables
$\widetilde{k}_3=k_3L_{\rm B}$, $ \widetilde{k}_\perp =k_\perp
L_{\rm B}$ are used.  %We introduced the Larmour radius $L_{\rm
%B}=(eB)^{-1/2}=(m^2b)^{-1/2}$, which  is much less than the
%Compton length $m^{-1}$ in our case.
We refer to Eq.
(\ref{universal}) as establishing a scaling regime that describes
the potential measured in inverse Larmour units as a universal
function of coordinates measured in Larmour units. It holds for
$|{\bf x}|\ll (2m)^{-1}$.

\begin{figure}
\includegraphics[width=0.47\textwidth]{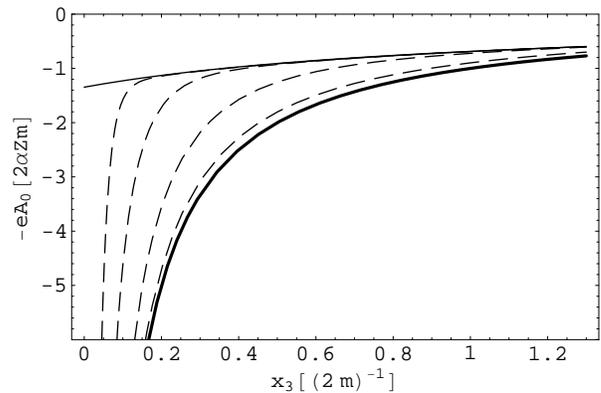}
\caption{\label{fig:1} Electron potential energy $-eA_0$ in the
field of a point charge $q=Ze$ plotted in units $2\alpha
Zm=Z\times$7.46~keV as a function of longitudinal coordinate $x_3$
in Compton half-lengths $(2m)^{-1}$ at $x_\perp =0$ for four
values of magnetic field (from left to right): $b=10^6,\; 10^5,\;
10^4,\; 10^3,\;b=B/B_0$,
 $B_0=4.4\times 10^{13}$ G (dashed lines). Solid
line is the fit $-1.4/(2mx_3+1.04)$. Bold line is the Coulomb law
$-1/2mx_3$.}
\end{figure}

\begin{figure}
\includegraphics[width=0.47\textwidth]{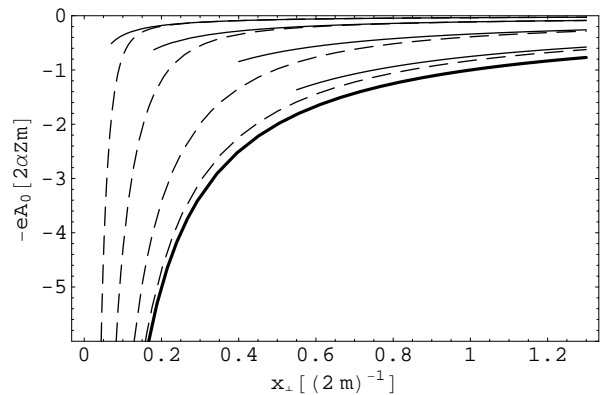}
 \caption{\label{fig:2} The same as in Fig.\ref{fig:1} in
function of transverse coordinate $x_\perp$ at $x_3 =0$. The
asymptotes at large $x_\perp$ values, $-1/2m x_\perp^\prime,$ are
presented by  solid lines.}
\end{figure}

The simple representation (\ref{universal}) can be further
simplified if $x_3$ or $x_\perp$ are large in the Larmour scale:
$|\widetilde{x}_3|\gg 1,$ or $|\widetilde{x}_\perp|\gg 1$ (but
remain small in the Compton scale). %, otherwise (\ref{universal})
%would become invalid).
In this case the integration in
(\ref{universal}) is restricted to the domain
$\widetilde{k}_\perp^2\ll 1$ where the exponential $\exp
(-\widetilde{k}_\perp^2/2)$ should be taken as unity. Then
(\ref{universal}) is reduced to the isotropic Yukawa law
\begin{eqnarray}\label{yuk}A_0({\bf x})\simeq \frac q{4\pi L_{\rm B}}\frac
{\exp\left(-{(2\alpha}/\pi)^{1/2}{\sqrt{\widetilde{x}_\perp^2+\widetilde{x}_3^2}}
\right)}{\sqrt{\widetilde{x}_\perp^2+\widetilde{x}_3^2}}.
\end{eqnarray} This can be established by tracing (\ref{universal})
back to (\ref{A0}) with
\begin{eqnarray}\label{mass}-\kappa_2(\infty,0)=\frac{2\alpha}{\pi L_{\rm
B}^2}=\frac{2\alpha b}{\pi}m^2=M^2\end{eqnarray} substituted for
$-\kappa_2(k_3^2, k_\perp^2)$ in the denominator. Here $M$ is the
"effective photon mass" $\;$noted in Ref. \cite{kuznets}. The
Yukawa law (\ref{yuk}) establishes the short-range character of
the static electromagnetic forces in the Larmour scale. Stress,
however, that the genuine photon mass understood as its rest
energy is always strictly equal to zero as a consequence of the
gauge invariance reflected in the approximation-independent
relation $\kappa_a(0,0)=0$ respected by (\ref{2}). Hence, the
potential, produced by a static charge, should be long-range for
sufficiently large distances. This is the case, indeed.  One can
see by inspecting the curves of Fig.1, computed using Eq.
(\ref{A0}) for $x_\perp=0$, that the scaling regime
(\ref{universal}), or (\ref{yuk}) does fail for sufficiently large
values of ${x}_3$: $\sim 0.1/(2m)=50L_{\rm B} $ for $b=10^6$,
$\sim 0.2/(2m)=33L_{\rm B}$ for $b=10^5$, $\sim 0.3/(2m)=15L_{B}$
for $b=10^4$ -  the larger, the smaller the field [for these
distances Eqs. (\ref{universal}) and (\ref{yuk}) are already the
same]. Starting with these values, the potential curves approach
their envelope, that can be fitted as $A_0(x_3,0)\simeq (q/4\pi)
1.41/(x_3+1.04/2m)$, unlike the scaling curves (\ref{universal}),
and (\ref{yuk}) that tend fast to zero. It is in this place that
the abruptly falling - short-range - potential turns into a slowly
decreasing - along the envelope curve - long-range potential. An
analogous change from the short- to long-range behavior is
observed in Fig. 2, where the electron energy in the field
(\ref{A0}) is plotted against  the transverse distance from the
charge at $x_3$=0. At small distances all the dashed curves in
Figs. \ref{fig:1} and \ref{fig:2} approach the thick solid Coulomb
curve in accord with (\ref{expand}).

 The
scaling regime (\ref{universal}), %in particular, the Yukawa regime
(\ref{yuk}), depends on the fact that the eigenvalue (\ref{2})
grows linearly with the magnetic field. If this linearity,
supposedly, retains in higher-loop approximations, one may
conjecture % claim the hypothesis
that the calculations of the
latter may be reduced to finding $\alpha^n$-corrections to the
mass (\ref{mass}). Potential (\ref{universal}) implies the
suppression of electrostatic force by the linearly growing term in
the denominator of (\ref{A0}) at large  distances in the Larmour
scale from the charge, but not close to the charge, where it has
the
same singularity as the Coulomb law. %, since the integral
%(\ref{universal}) diverges in the point $x_\perp=x_3=0$ in the
%same way as if $\alpha$ in it were set equal to zero. % (see,
%however, the reservation \cite{footnote}). %We should note, however, that
%when we are so close to the charge, $x_3 < L_{\rm B}$, that the
%regime of external field dominance ceases, the expression
%(\ref{2}) that we were using literally in the above analysis is
%itself no longer valid, and the genuine asymptotic behavior of the
%potential is not the leading Coulomb term in (\ref{expand}), but
%its known modification due to the vacuum polarization with no
%external field \cite{blp}. But this occurs at too small distances,
%$m^{-1}\exp(-1/\alpha),$ beyond interest.

For larger distances $x_3$, $x_\perp$, not small in the Compton
scale, one may expect that only integration over small $k_3$,
$k_\perp$ is important in (\ref{A0}) (for a more thorough analysis
see \cite{SUPRD2}). In this limit (\ref{2}) behaves as
$\kappa_2(k_3^2, 0)\simeq -\frac{\alpha b}{3\pi}k_3^2.$ With this
substitution the integral in (\ref{A0}) is
\bee\label{llargex}A_0(x_3,x_\perp)\simeq \frac 1{4\pi}\frac
{q}{\sqrt{(x^\prime_\perp)^2+x_3^2}},\eend where $ x^\prime_\perp
=\beta x_\perp ,\;\beta\equiv\left(1+\alpha
b/3\pi\right)^{1/2},~~x^\prime_\perp>x_\perp.$ %A more thorough
%analysis \cite{SUPRD2} demonstrates that approximation
%(\ref{llargex}) is valid, if either $x_\perp\gg L_{B},$ or
%$x_3\gg(2m)^{-1}$, in the latter case with the accuracy to
%exp$(-2m|x_3|)$.
For small $\alpha b/3\pi$, Eq. (\ref{llargex})
coincides with the result of \cite{loskutov2}.

Eq. (\ref{llargex}) is an "anisotropic Coulomb law", according to
which the attraction force decreases with distance from the source
along the transverse direction faster than along the magnetic
field, but remains long-range.  In accord with (\ref{llargex}),
the curves $A_0(x_3,0)$  in Fig.\ref{fig:1} all approach at large
distances $|x_3|$
 the Coulomb law $q/(4\pi |x_3|),$ whereas each curve $A_0(0,x_\perp)$
 in Fig. \ref{fig:2} reaches at large $x_\perp$
the asymptote $A_0(0,x_\perp)=q/(4\pi x'_\perp) = A_0^{\rm
C}(0,x_\perp)/\beta$, different for each field, {\it i.e.,} the
potential is anisotropic. Again, the same as for short distances
considered above, we face - now anisotropic - suppression of the
Coulomb force due to the linearly growing term in (\ref{2}). The
equipotential surface is an ellipsoid stretched along the magnetic
field. The electric field of the charge  ${\bf E}=-{\bf
\nabla}A_0(x_3,x_\perp)$ is a vector with the components
$(q/2\pi)(x_3^2+\beta^2x_\perp^2)^{-3/2}(x_3,\beta^2{\bf
x}_\perp)$. It is not directed towards the charge, but makes an
angle $\phi$ with the radius-vector $\bf r$,
cos$\phi=(x_3^2+\beta^2x_\perp^2)(x_3^2+\beta^4x_\perp^2)^{-1/2}
(x_3^2+x_\perp^2)^{-1/2}.$ In the limit of infinite magnetic
field, $\beta=\infty$, the electric field of the point charge is
directed normally to the axis $x_3$.

The regime (\ref{llargex}) corresponds to the approximation, where
only quadratic terms in powers of the photon momentum are kept in
$\kappa_2$. Within this scope the dielectric permeability of the
vacuum is independent of the frequency, and the refractive index
depends only upon the angle in the space \cite{shabad2004}.

%To exemplify an implementation of the modified Coulomb potential,
Consider an impact of the Coulomb potential modification on the
nonrelativistic hydrogenlike atom ground state energy. The wave
function $\Psi(x_3)$ is subject to one-dimensional Schr$\ddot{\rm
o}$dinger equation \cite{{elliott}} with respect to $x_3,$ valid
in the region $|x_3|>L_{\rm B}$. The transverse coordinate
$x_\perp$ in the argument of the potential is replaced by the
momentum of the transverse center-of-mass motion \cite{ShUs} and
should be set equal to zero for the ground state.

For the fields in the range $1\ll b\ll 2\pi/\alpha\sim 10^3$ the
correction (\ref{difference2}) to the Coulomb potential may be
treated as perturbation. Keeping only the first-order term in
$\kappa_2$ in (\ref{difference2}) we may derive the magnetic
analog of the Uehling potential \cite{blp}. At $x_\perp =0$ it is
\bee\label{uehling2}\Delta A_0(x_3,0)\simeq \frac{q\alpha b
m}{8\pi^2}\int_0^{\pi/2}
\rme^{-(2m|x_3|/\cos\phi)}\cos^2\phi~\rmd\phi.\eend This
correction is of the order of $\alpha b$, {\it i.e.,} is much
larger than $\alpha$. Eq. (\ref{uehling2}) implies $C\simeq\alpha
b/16$. Calculating its matrix element  with Loudon's
\cite{elliott} wave functions we obtain the positive correction
($q=Ze$) \bee\label{perturbed2} \frac{2Z^2\alpha^3 b m}{3\pi
}\ln\frac b{4\alpha^2}= Z^2b~\left(\frac{\ln
b}{8.454}+1\right)\times 0.356~{\rm eV}\eend to the negative
Loudon-Elliott ground state
energy\bee\label{LE} E_0 %=-\frac {2Z^2}{ma_{\rm
%B}^2}\ln^2\frac{2L_{\rm B}}{a_{\rm B}}
=-2Z^2\alpha^2m\ln^2\frac{\sqrt{b}}{2\alpha}.\eend %Here $a_{\rm
%B}=(m\alpha)^{-1}$ is the Bohr radius.
Eq. (\ref{perturbed2}) may
be used for $Z \leq 6$. For instance, for $Z=1$ and $b= 200$ it
makes about 100 eV, while $E_0\approx -2.5$ keV. Although the
proton has a finite size $R\sim 10^{-13}{\rm cm}$, the Coulomb
$1/|x_3|$ part of the potential (\ref{expand}) remains the same
within the definition range of the
 Schr$\ddot{\rm o}$dinger equation, since $R < L_{\rm B}$ within
the range of $b$ considered in this paragraph (moreover, $R \ll
L_{\rm B})$. On the contrary, the vacuum-polarization part does
depend on $R,$ like it does in no-magnetic-field case \cite{mohr}.
However, the finite-size correction to (\ref{perturbed2}) is
$\sim(R/L_{\rm B})^2$ to be neglected within the present scope of
accuracy.

The singularity  $1/|x_3|$ of the Coulomb potential in the origin
is known to lead to the energy spectrum unbounded from below
\cite{elliott}: as this singularity is cut off at the Larmour
length, $L_{\rm B}=(m\sqrt{b})^{-1}$, the ground state energy
(\ref{LE}) tends to $-\infty$  with the growth of the magnetic
field, when $L_{\rm B}\rightarrow 0$. This feature is cured by the
vacuum polarization. As $b\rightarrow\infty$ the region in Fig.1
between the potential curve and the abscissa below the point
$-eV\approx -1.4\times 2\alpha Zm$, where it is crossed by the
envelope, becomes infinitely deep and thin. The area
$S=(4\pi/q)\int_{L_{\rm B}}^{\overline{x}_3} A_0(x_3,0)\rmd x_3$
calculated with expression (\ref{expand}), where
${\overline{x}_3}$ is found from the equation
$A_0(\overline{x}_3,0)=V$, has a finite limit if and only if $C$
grows proportionally to $1/L_{\rm B}$. This is the case: when
calculated following Eq. (\ref{universal}), $C\simeq
0.9594\sqrt{\alpha b/2\pi}$. Thus, the dominating behavior of the
modified Coulomb potential in the origin becomes the
$\delta$-function. Combining it with the fit for the envelope we
come to the limiting form of the electron potential energy at
$b=\infty$ (here $|x_3|$ does not exceed a few $(2m)^{-1}$)
\bee\label{limiting}\hspace{-0.5cm}-eA_0(x_3,0)=-2\alpha Z
\left(S~\delta(x_3)+\frac{1.4m}{2m|x_3|+1.04}\right),\eend where
$S=\ln (\sqrt{\pi/2\alpha}/0.96)-1+0.96\sqrt{2\alpha/\pi}=1.79.$ A
more rigorous calculation done with the use of eqs.
(\ref{universal}) or (\ref{yuk}) for the short-range part of the
potential instead of (\ref{expand}) results in $S= 2.18\approx
{\rm -Ei}(-(2\alpha/\pi)^{1/2}),$ Ei is the exponential integral.
With the $1/|x|$-singularity replaced by the $\delta$-function,
the ground energy level is certainly finite. Applying the formula
for the ground energy $-2m(\int_{L_{\rm B}}^{a_{\rm B}}
A_0(x_3,0)\rmd x_3)^2$ valid in a shallow well potential
\cite{QM}, with eq. (\ref{limiting}) integrated up to the point
$x_3=2.6/2m$, where the fitted curve crosses the Coulomb law, and
the latter  taken as the integrand for larger $x_3$ up to the Bohr
radius $a_{\rm B}=(m\alpha)^{-1}$, we estimate the {\em finite}
limiting value for the ground energy as \bee\label{limenergy}
E_{\rm lim}=-2mZ^2\alpha^2~73.8=-Z^2\times 4 ~{\rm keV}.\eend  The
Loudon-Elliott energy (\ref{LE}) would overrun the limiting energy
(\ref{limenergy}) already for the magnetic field as large as
$b=6600$, when yet $R\ll L_{\rm B}$. The ground level reaches 92\%
of its limiting value for $b=5\times 10^4$. After the magnetic
field reaches the value $b=1.5\times 10^5$, when $R$ and $L_{\rm
B}$ equalize, the Coulomb potential is cut off at the proton size,
$x_3=R$. Setting $L_{\rm B}=R$ in (\ref{LE}) we would get the
minimum value for the Loudon-Elliott energy ($Z=1$) to be $-5.6$
keV, which is essentially lower than (\ref{limenergy}).

The significant modification of the Coulomb potential of an
electric charge by the vacuum polarization in external constant
magnetic field $B\gg B_0$, shown to eliminate the unboundedness
from below of the nonrelativistic hydrogen spectrum, is apt of
having more implementation as far as other situations where
electrostatic fields are important are concerned, {\em e.g.}
properties of matter on surfaces of extremely magnetized neutron
stars.

This work was supported by the Russian Foundation for Basic
Research (project no 05-02-17217) and the President of Russia
Programme (LSS-4401.2006.2), as well as by the Israel Science
Foundation of the Israel Academy of Sciences and Humanities.

\vfill\eject
\newpage

\begin{thebibliography}{20}
\bibitem{footnote} Except for far ultraviolet region,
where $\kappa_2$, replaced by its  value with no magnetic field,
grows  like $(\alpha/3\pi)k^2\ln (k^2/m^2)$, which changes the
potential to the extent comparable with that considered in this
paper only for distances as small as $(1/m)\exp (-3\pi/2\alpha)$
\cite{blp}
\end{thebibliography}
\end{document}